\documentclass[prl,twocolumn,superscriptaddress]{revtex4}
\usepackage{amsmath}
\usepackage{amssymb}
\usepackage{amsthm}
\usepackage{amsfonts}
\usepackage{listings}
\usepackage{enumerate}
\usepackage{latexsym}
\usepackage{color}
\usepackage{xcolor}
\usepackage{bm}
\usepackage{hyperref}
\hypersetup{
 pdfnewwindow=true, colorlinks=true,
 linkcolor=blue, anchorcolor=blue,
 citecolor=blue, filecolor=blue,
 menucolor=blue, urlcolor=blue}

\newcommand{\beginsupplement}{%
        \setcounter{table}{0}
        \renewcommand{\thetable}{S\arabic{table}}%
        \setcounter{figure}{0}
        \renewcommand{\thefigure}{S\arabic{figure}}%
     }

\usepackage{psfrag}

\usepackage{bm}
\usepackage{graphicx}
\usepackage{subfigure}

\RequirePackage[normalem]{ulem} %DIF PREAMBLE
\RequirePackage{color}\definecolor{RED}{rgb}{1,0,0}\definecolor{BLUE}{rgb}{0,0,1} %DIF PREAMBLE
\newcommand{\bra}[1]{\left\langle#1\right|}
\newcommand{\ket}[1]{\left|#1\right\rangle}

\newcommand{\up}{\uparrow}
\newcommand{\dw}{\downarrow}

\def\ie{{\it i.e.},\ }

\input{epsf}

\begin{document}

\tolerance 10000

\newcommand{\vk}{{\bf k}}

\draft

\title{Electronic structures and topological properties in nickelates $Ln_{n+1}$Ni$_n$O$_{2n+2}$}

%in English titles articles and words like to, on, at etc are always spelled with small letters
\author{Jiacheng Gao}
\affiliation{Beijing National Laboratory for Condensed Matter
Physics, Institute of Physics, Chinese Academy of Sciences, Beijing
100190, China}
\affiliation{School of Physical Sciences, University of Chinese Academy of Sciences, Beijing 100190, China}
\author{Shiyu Peng}
\affiliation{Beijing National Laboratory for Condensed Matter
Physics, Institute of Physics, Chinese Academy of Sciences, Beijing
100190, China}
\affiliation{School of Physical Sciences, University of Chinese Academy of Sciences, Beijing 100190, China}
\author{Zhijun Wang}
\email{wzj@iphy.ac.cn}
\affiliation{Beijing National Laboratory for Condensed Matter
Physics, Institute of Physics, Chinese Academy of Sciences, Beijing
100190, China}
\affiliation{School of Physical Sciences, University of Chinese Academy of Sciences, Beijing 100190, China}

\author{Chen Fang}
\email{cfang@iphy.ac.cn}
\affiliation{Beijing National Laboratory for Condensed Matter
Physics, Institute of Physics, Chinese Academy of Sciences, Beijing
100190, China}
\affiliation{School of Physical Sciences, University of Chinese Academy of Sciences, Beijing 100190, China}

\author{Hongming Weng}
\email{hmweng@iphy.ac.cn}
\affiliation{Beijing National Laboratory for Condensed Matter
Physics, Institute of Physics, Chinese Academy of Sciences, Beijing
100190, China}
\affiliation{School of Physical Sciences, University of Chinese Academy of Sciences, Beijing 100190, China}
\affiliation{Songshan Lake Materials Laboratory, Dongguan, Guangdong 523808, China }
\affiliation{CAS Centre for Excellence in Topological Quantum Computation, Beijing, China}
\affiliation{Physical Science Laboratory, Huairou National Comprehensive Science Center, Beijing, China}

\begin{abstract}
After the significant discovery of the hole-doped nickelate compound Nd$_{0.8}$Sr$_{0.2}$NiO$_2$, an analysis of the electronic structure, orbital components, Fermi surfaces and band topology could be helpful to understand the mechanism of its superconductivity. Based on the first-principles calculations, we find that Ni $3d_{x^2-y^2}$ states contribute the largest Fermi surface. $Ln~5d_{3z^2-r^2}$ states form an electron pocket at $\Gamma$, while $5d_{xy}$ states form a relatively bigger electron pocket at A. These Fermi surfaces and symmetry characteristics can be reproduced by our two-band model, which consists of two elementary band representations: $B_{1g}@1a~\oplus~A_{1g}@1b$. 
We find that there is a band inversion near A, giving rise to a pair of Dirac points along A--M below the Fermi level once including spin-orbit coupling.
Furthermore, we have performed the LDA+Gutzwiller calculations to treat the strong correlation effect of Ni 3d orbitals. In particular, the bandwidth of $3d_{x^2-y^2}$ has been renormalized largely.   After the renormalization of the correlated bands, 
the Ni $3d_{xy}$ states and the Dirac points become very close to the Fermi level.
Thus, a hole pocket at A could be introduced by hole doping, which may be related to the observed sign change of Hall coefficient. 
By introducing an additional Ni $3d_{xy}$ orbital, the hole-pocket band and the band inversion can be captured in our modified model. Besides, the nontrivial band topology in the ferromagnetic two-layer compound La$_3$Ni$_2$O$_6$ is discussed and the band inversion is associated with Ni $3d_{x^2-y^2}$ and La $5d_{xy}$ orbitals.
\end{abstract}

\maketitle
\section{Introduction}

After the discovery of high-T$_c$ superconductivity in the cuprates\cite{bednorz1986possible,wu1987superconductivity}, mixed-valent nicklates with similar crystal and electronic configurations as cuprates have been attracting lots of attention~\cite{anisimov1999electronic}. In particular, the configuration of Ni$^+$ in infinite-layer nickelates $Ln$NiO$_2$ ($Ln$=La, Nd, Pr) is almost identical to that of Cu$^{++}$ in the parent compounds of cuprates. Although much effort has been devoted along this direction in the past two decades, it remains elusive for the possible superconductivity in mixed-valent nicklates. Until very recently, the superconductivity with T$_c=9\sim 15$K was discovered in hole-doped Nd$_{0.8}$Sr$_{0.2}$NiO$_2$ for the first time\cite{NickelNature}. For the parent compound NdNiO$_2$, previous studies have established several experimental facts that are distinct from the parent compound of cuprates. First, no long-range magnetic order is observed experimentally\cite{hayward1999sodium,hayward2003synthesis}, while an antiferromagnetically ordered state is formed in the cuprates\cite{motoyama2007spin}. Second, NdNiO$_2$ exhibits a metallic behavior above 50K\cite{NickelNature}, while the parent cuprates are Mott insulators\cite{lee2006doping}. Third, the superconductivity (so far) is only found in the hole-doped Nd$_{0.8}$Sr$_{0.2}$NiO$_2$, while it is found in the electron-doped cuprate Sr$_{1-x}$La$_x$CuO$_2$ in the same structure\cite{smith1991electron}. These experimental facts indicate that the ground state of the parent nickelates could have significant difference from the cuprates. The analysis of their electronic band structures, orbital components, Fermi surfaces, and the band topology are wanted. In addition, a minimal-band effective model is very helpful to further understand the mechanism of superconductivity.

\begin{figure}[!tb]
\centering
\includegraphics[width=7.5 cm]{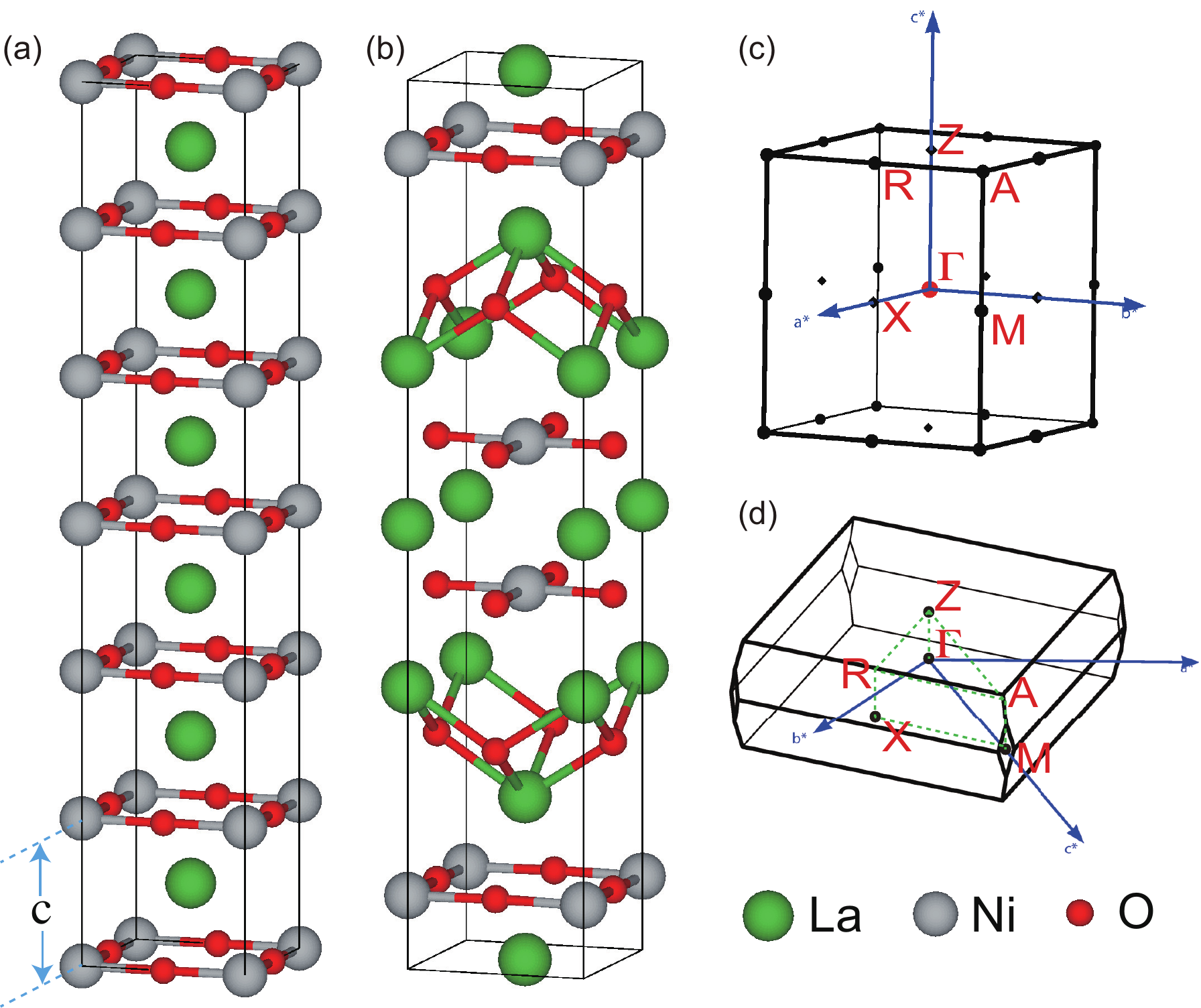}
\caption{(Color online)
Crystal structures and Brillouin zones.
The crystals of infinite-layer compound LaNiO$_2$ ($P4/mmm$) and two-layer compound La$_3$Ni$_2$O$_6$ ($I4/mmm$) are presented in (a) and (b), respectively. (a) contains six unit cells ($c$ is the lattice parameter in $z$ direction). The primitive reciprocal lattice vectors and high-symmetry $k$-points are indicated in the first Brillouin zones of LaNiO$_2$ (c) and La$_3$Ni$_2$O$_6$ (d).
} \label{fig:1}
\end{figure}

\begin{figure*}[!t]
\centering
\includegraphics[width=16 cm]{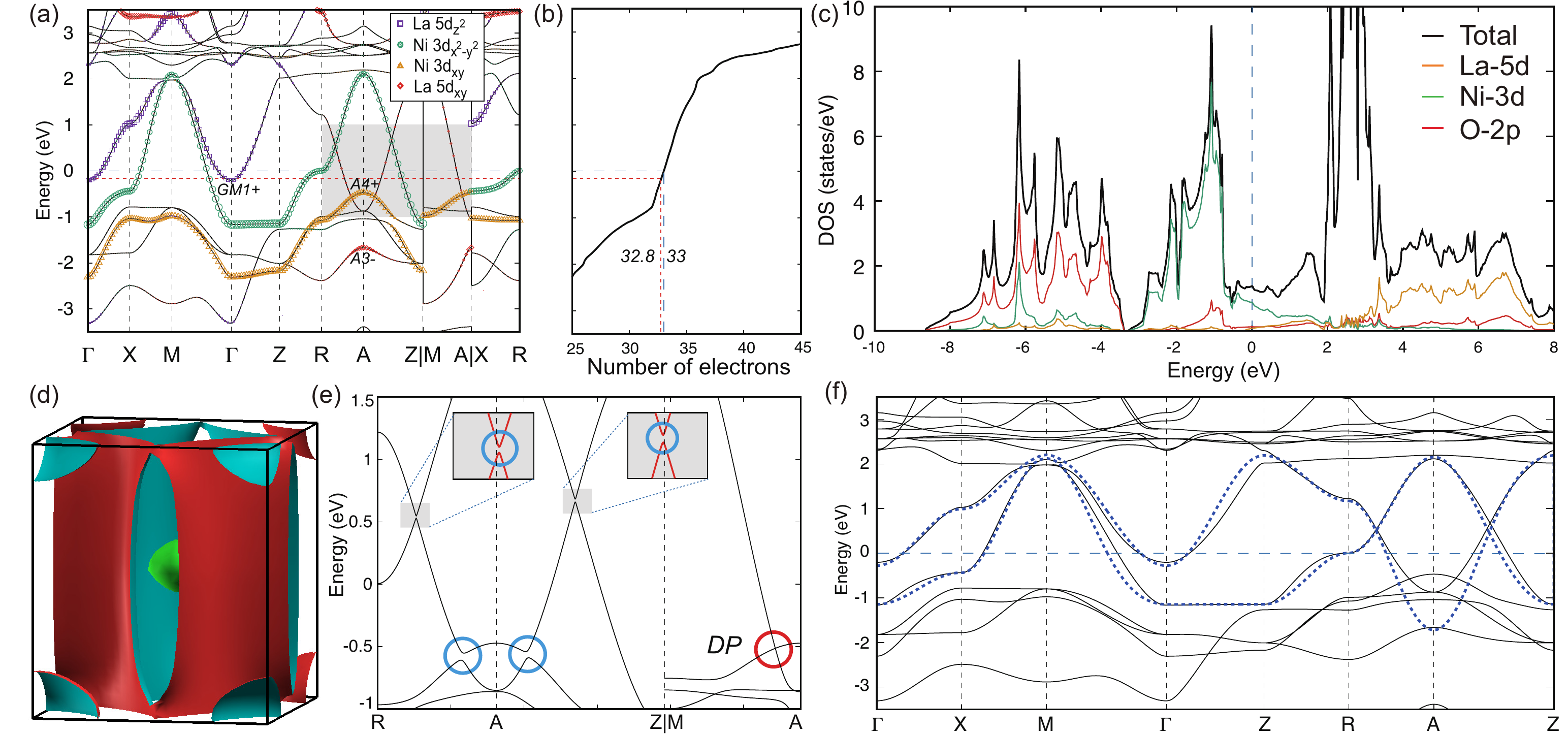}
\caption{(Color online)
(a) The band structure of LaNiO$_2$ without SOC. The weights of La $5d_{z^2}$, Ni $3d_{x^2-y^2}$, Ni $3d_{xy}$ and La $5d_{xy}$ states are indicated by the size of the purple squares, green circles, yellow triangles and red diamonds, respectively. The La $5d_{z^2}$ band at $\Gamma$ is labeled by $GM1+$, while the Ni $3d_{xy}$ and La $5d_{xy}$ bands at A are labeled by $A4+$ and $A3-$, respectively.
The blue dashed line represents the Fermi level $E_F$ and the red dashed line represents the estimated chemical potential in the hole doped Nd$_{0.8}$Sr$_{0.2}$NiO$_2$. The total number of electrons as a function of chemical potential is plotted in (b). The partial density of states (DOS) is given in (c).  The Fermi surfaces of LaNiO$_2$ are shown in (d). In the band structure with SOC (e), the crossings in the shadowed area of (a) are gapped except the Dirac point (DP) along M--A. The bands of our two-band model are shown as blue dashed lines in (d).
} \label{fig:2}
\end{figure*}

In this work, we have performed detailed first-principles calculations within the framework of density functional theory (DFT/LDA). The obtained band structure of the parent(un-doped) compound NdNiO$_2$ is similar to that of LaNiO$_2$ reported previously \cite{lee2004infinite} and PrNiO$_2$ (Nd-$4f$ and Pr-$4f$ orbitals are treated as core states.). In this article, LaNiO$_2$ is taken as a representative and the band structures of NdNiO$_2$ and PrNiO$_2$ are presented in the Supplemental Material (SM). In the un-doped case, there are three bands intersecting the Fermi level (E$_F$), mainly from Ni-$3d_{x^2-y^2}$, La-$5d_{xy}$ and La-$5d_{3z^2-r^2}$ orbitals. They form a large cylinder-like electron pocket (EP) surrounding the $\Gamma$--Z line, a sphere-like EP at A, and a comparatively smaller sphere-like EP at $\Gamma$, respectively. 
These Fermi surfaces and symmetries can be reproduced by our two-band model, which consists of two elementary band representations (EBRs): $B_{1g}@1a~\oplus~A_{1g}@1b$. The EBR of $B_{1g}@1a$ refers to $3d_{x^2-y^2}$ orbital at Wyckoff site $1a$, while the EBR of $A_{1g}@1b$ refers to $A_{1g}$ orbital at Wyckoff site $1b$, where no atoms sit.

%In the hole-doped case, the small EP at $\Gamma$ is removed, while another hole pocket (HP) is growing at A, which is mainly from Ni $3d_{xy}$ orbital. This HP may be responsible for the sign change of Hall coefficient in the experiment~\cite{NickelNature}.
We find a band inversion near A happens between the Ni $3d_{xy}$ states and $Ln~5d_{xy}$ states (The effect of Coulomb interaction $U$ is discussed in the SM). 
With small $U$ and spin-orbital coupling (SOC), it gives rise to a pair of Dirac points along A--M. After considering the renormalization of Ni $3d$ bands, the Dirac point becomes very close to the charge neutrality level ($E_F$) and accessible by hole doping. As a result, a hole pocket (HP) may emerge at A, which may be responsible for the sign change of Hall coefficient in the experiment~\cite{NickelNature}. By introducing an additional Ni $3d_{xy}$ orbital, the hole-pocket band and the band inversion can be captured in the modified model. Besides, the nontrivial band topology in the ferromagnetic two-layer ($n=2$) compound La$_3$Ni$_2$O$_6$ is discussed and a band inversion happens between Ni $3d_{x^2-y^2}$ and La $5d_{xy}$ orbitals.

\section{Crystal structure and Methodology}
The parent compound LaNiO$_2$ can be obtained from the perovskites LaNiO$_3$ by removing the apical oxygens, as shown in Fig.~\ref{fig:1}(a). Consequently, it has a tetragonal lattice, and has the same planes as the cuprate superconductors with Ni$^+$ instead of Cu$^{++}$ ions. Similarly, the two-layer nickelate La$_3$Ni$_2$O$_6$ [in Fig.~\ref{fig:1}(b)] can be produced \cite{PRL.102.046405} from the two-layer perovskite La$_3$Ni$_2$O$_7$.
We performed the first-principles calculations with VASP package \cite{KRESSE199615,vasp} based on the density functional theory with the projector augmented wave (PAW) method \cite{paw1,paw2}.
 %were employed in our first-principles calculations.
The generalized gradient approximation (GGA) with exchange-correlation functional of Perdew, Burke and Ernzerhof (PBE) for  the exchange-correlation functional \cite{pbe} were employed. 
The kinetic energy cutoff  was set to 500 eV for the plane wave basis. 
A $10\times 10\times 10$ $k$-mesh in self-consistent process for Brillouin zone (BZ) sampling was adopted. The experimental lattice parameters of LaNiO$_2$ and La$_3$Ni$_2$O$_6$ are employed~\cite{NickelNature,PRL.102.046405}.
%The lattice parameters of LaNiO$_2$ and La$_3$Ni$_2$O$_6$ from Refs. \cite{NickelNature} and ~\cite{PRL.102.046405} respectively are used in this study.

\section{Results and discussions} 

\subsection{Band structure and Density of states}
We have first performed the first-principles calculations on LaNiO$_2$ without SOC. The band structure without SOC is presented in Fig.~\ref{fig:2}(a) and the total density of states (DOS) is plotted accordingly.
The blue dashed horizontal line corresponds to the charge neutrality level of the un-doped compound LaNiO$_2$. Based on the total number of electons in Fig.~\ref{fig:2}(b), the red-colored dashed line is the theoretically estimated chemical potential for the 20\% hole-doped superconductivity Nd$_{0.8}$Sr$_{0.2}$NiO$_2$.
Moreover, the partial DOSs are also computed for O $2p$, Ni $3d$, and La $5d$ orbitals, respectively. Since the main quantum numbers of different atom's orbitals are distinct, we call them $2p$, $3d$, and $5d$ orbitals (states) for short in the following discussion.
From the plotted partial DOS in Fig.~\ref{fig:2}(c), we have noticed that $2p$ states are mainly located from $-10$ eV to $ -3.5$ eV below E$_F$, while $3d$ states are around E$_F$, from $-3.5$ eV and $1.5$ eV. The situation is much different from the situation in copper-based superconductors, where O $2p$ states are slightly below $E_F$ and hybridize strongly with Cu $3d$ states~\cite{lee2004infinite}.
In addition, from the orbital-weighted fat bands in Fig.~\ref{fig:2}(a), we notice that there are $5d_{3z^2-r^2}$ states at $\Gamma$ and $5d_{xy}$ states at $A$ around E$_F$, suggesting that the 5d states are more extended, compared to Ca $3d$ states in CaCuO$_2$~\cite{wu1999electronic}.

\subsection{Evolution of Fermi surfaces}
At the charge neutrality level, we find that there are three bands crossing $E_F$, which are mainly from $3d_{x^2-y^2}$, $5d_{z^2}$ and $5d_{xy}$ orbitals, respectively. the weights of these orbitals are depicted by the size of different symbols in Fig.~\ref{fig:2}(a). Therefore, three EPs are formed as shown in Fig.~\ref{fig:2}(d): i) $3d_{x^2-y^2}$ orbital forms the largest electron pocket around the $\Gamma$ and $Z$ points, which has a strong 2D feature; ii) the second larger one is nearly a sphere around A, formed by $5d_{xy}$ orbital; iii) the smallest one is a sphere around $\Gamma$, formed by $5d_{3z^2-r^2}$ orbital (which is also hybridized with Ni $3d_{3z^2-r^2}$ orbital in the DFT calculations, yet we still call it $5d_{3z^2-r^2}$ for simplicity). 

In the hole-doped superconductor Nd$_{0.8}$Sr$_{0.2}$NiO$_2$, the estimated chemical potential of the 20\% Sr-doped level is denoted by a red-colored dashed line, which corresponds to 32.8 electrons per unit cell (The charge neutrality level corresponds to 33 electrons). 
Needless to say that, all the electron pockets become smaller with hole-doping. In particular,  the $5d_{3z^2-r^2}$-orbital-formed $\Gamma$-centered EP is about to be removed. On the other hand, the states from $3d_{xy}$ orbital becomes closer to the chemical potential, especially in the vicinity of A point.
%The energy band of Ni $3d_{xy}$ orbital at A is higher than that at M, indicating that the interlayer coupling between them is substantial.

\subsection{Band inversion and Dirac points}
The band crossings along Z--R, R--A, M--A are protected by $m_{110}$, $m_{001}$ and C$_{4z}$, respectively. 
After considering SOC, the band crossings open small gaps along the mirror protected R--A and A--Z lines, but remain gapless along the C$_{4z}$-invariant line M--A, as shown in Fig.~\ref{fig:2}(e). 
The gapless Dirac points along M--A [highlighted in the red circle in Fig.~\ref{fig:2}(e)] are protected by $C_{4v}$ symmetry. Namely, the two doubly-degenerate bands belong to different 2D irreducible representations (irreps) of $C_{4v}$ double group. 
In our LDA+U calculations, we find that the band inversion is sensitive to the value of Coulomb interaction $U$, as we have shown in the SM. 
%To get more information about Dirac points and construct a model to capture the anti-crossing band structure, we need to know the band representation under little groups, which is govern by compatibility relation and topological quantum chemistry.

\begin{table}[!t]
\caption{
The upper rows give the irreps for lowest six bands at in Fig. 2(a) at the maximal high-symmetry points in SG 123. The irreps are given in ascending energy order. The notation of $Zm(n)$ implies the irrep $m$ at the $Z$ point with the degeneracy of $n$. All the irrep notations are listed on the Bilbao website: www.cryst.ehu.es/cgi-bin/cryst/programs/bandrep.pl.
The lower rows give the elementary band representation (EBRs), labeled as $\rho @q$. Here, $\rho$ indicates the irrep supported by the orbital(s), while $q$ stands for the Wyckoff site, where the orbital(s) sit. The green irreps indicate that those energy bands are at least 1.0 eV above E$_F$.
}\label{tab:irreps}
\begin{tabular}{c|cccccc}
\hline
\hline
& A & $\Gamma$ & M & Z & R & X \\
\hline
DFT&{\color{blue} $A3-$(1)} & GM1+(1) & M1+(1) & Z4+(1) & R4+(1) & X1+(1) \\
bands& A1+(1) & GM4+(1) & M4+(1) & Z5+(2) & R1+(1) & X4+(1) \\
& A5+(2) & GM5+(2) & M5+(2) &        & R2+(1) & X2+(1) \\
&        &         &        & Z1+(1) & R3+(1) & X3+(1) \\
&   ;    &   ;     &   ;    &   ;    &   ;    &    ;   \\
&{\color{blue} $A4+$(1)} & GM2+(1) &  & Z2+(1) & R1+(1) & X1+(1) \\
%& A2+(1) & GM1+(1) &        & Z1+(1) & R2-(1) & X4-(1) \\
& A2+(1) & GM1+(1) &  {\color{gray!90} M$5-$(2)} &{\color{gray!90}  Z1+(1)} &{\color{gray!90}  R$2-$(1) }& {\color{gray!90} X$4-$(1) }\\
\hline
\hline
EBRs &&&&&&\\
\hline
$A_{1g}@1a$ &A1+&GM1+&M1+&Z1+&R1+&X1+\\
\hline
$B_{1g}@1a$ &A2+&GM2+&M2+&Z2+&R1+&X1+\\
\hline
$B_{2g}@1a$ &A4+&GM4+&M4+&Z4+&R2+&X2+\\
\hline
$E_{g}@1a$ &A5+&GM5+&M5+&Z5+&R3+&X3+\\
        &   &    &   &   &R4+&X4+\\
\hline
$A_{1g}@1d$ &A$2-$&GM1+&M4+&Z$3-$&R3+&X$4-$\\
\hline
%$B_{1g}@1d$ &A1-&GM2+&M3+&Z4-&R3+&X4-\\
%\hline
$B_{2g}@1d$ &{\color{blue} $A3-$}&GM4+&M1+&Z$2-$&R4+&X$3-$\\
\hline
$A_{1g}@1b$ &A$3-$&GM1+&M1+&Z$3-$&R$2-$&X1+\\
\hline
\hline
\end{tabular}
\end{table}

% \begin{table*}[t]
% \caption{
% The hopping parameters for the three-band Hamiltonian, from which one can also found the two-band model parameters
% }\label{tab:parawzj}
% \begin{tabular}{cc|cc|ccp{0.1cm}|p{0.1cm}cc|ccc|ccc|cc}
% \hline
% \hline
% \multicolumn{7}{c|}{parameters for the two-band model}&
% \multicolumn{10}{c}{parameters for the three-band model} \\
% \hline
%  $t_{11}^{(0,0,0)}$ & 2.7319  & $t_{22}^{(0,0,0)}$ & 0.1470 & $t_{21}^{(0,0,0)}$ & 0.0200 &&  &$t_{11}^{(0,0,0)}$ & 2.7319 & & $t_{22}^{(0,0,0)}$ & 0.1470 & & $t_{33}^{(0,0,0)}$  & 2.9955& $t_{21}^{(0,0,0)}$ & 0.0200 \\
%  $t_{11}^{(1,0,0)}$ & -0.4125 & $t_{22}^{(1,0,0)}$ & 0.0913& $t_{21}^{(1,0,0)}$ & 0.0100  &&&$t_{11}^{(1,0,0)}$ & -0.4125 & & $t_{22}^{(1,0,0)}$ & 0.0913 & & $t_{33}^{(1,0,0)}$ &-0.20125& $t_{21}^{(1,0,0)}$ & 0.0100\\
%  $t_{11}^{(0,0,1)}$ & -0.0538 & $t_{22}^{(0,0,1)}$ & 0.0650&& & &&$t_{11}^{(0,0,1)}$ & -0.0538 & & $t_{22}^{(0,0,1)}$ & 0.0650 & & $t_{33}^{(0,0,1)}$ &-0.0450& $t_{31}^{(0,1,1)}$ & 0.0200 \\
%  $t_{11}^{(1,1,0)}$ & 0.0894  & $t_{22}^{(1,1,0)}$ & -0.0606&&&&&$t_{11}^{(1,1,0)}$ & 0.0894  & & $t_{22}^{(1,1,0)}$ & -0.0606 & & $t_{33}^{(1,1,0)}$& 0.06375& $t_{32}^{(1,0,0)}$ & 0.0100\\
%  $t_{11}^{(1,0,1)}$ & 0.0000  & $t_{22}^{(1,0,1)}$ & 0.1988 &&& & &$t_{11}^{(1,0,1)}$ & 0.0000 & & $t_{22}^{(1,0,1)}$ & 0.1988 & & $t_{33}^{(1,0,1)}$  & 0.0156& &\\
%  $t_{11}^{(1,1,1)}$ & 0.0134  & $t_{22}^{(1,1,1)}$ & 0.0281 &&& & &$t_{11}^{(1,1,1)}$ & 0.0134 & & $t_{22}^{(1,1,1)}$ & 0.0281 & & $t_{33}^{(1,1,1)}$  &-0.0069& &\\
% \hline
% \hline
% \end{tabular}
% \end{table*}

\begin{table*}[t]
\caption{
The hopping parameters for the two-band and three-band Hamiltonian.
}\label{tab:parawzj}
\begin{tabular}{cc|cc|ccp{0.1cm}|p{0.1cm}cc|ccc|ccc|cc}
\hline
\hline
\multicolumn{7}{c|}{parameters for the two-band model}&
\multicolumn{10}{c}{parameters for the three-band model} \\
\hline
 $t_{11}^{(0,0,0)}$ & 0.1470 & $t_{22}^{(0,0,0)}$ & 0.8318 & $t_{21}^{(1,0,0)}$ & 0.0098 &&  &$t_{11}^{(0,0,0)}$ & 0.0100 & & $t_{22}^{(0,0,0)}$ & 0.8280 & & $t_{33}^{(0,0,0)}$ &-0.7960 & $t_{21}^{(1,0,0)}$ & 0.0098 \\
 $t_{11}^{(1,0,0)}$ &-0.4125 & $t_{22}^{(1,0,0)}$ & 0.0913 &                    &        &&  &$t_{11}^{(1,0,0)}$ & 0.0042 & & $t_{22}^{(1,0,0)}$ &-0.0510 & & $t_{33}^{(1,0,0)}$ &-0.1655 & $t_{31}^{(1,0,0)}$ &-0.0132 \\
 $t_{11}^{(0,0,1)}$ &-0.0538 & $t_{22}^{(0,0,1)}$ & 0.0650 &                    &        &&  &$t_{11}^{(0,0,1)}$ & 0.0378 & & $t_{22}^{(0,0,1)}$ &-0.0079 & & $t_{33}^{(0,0,1)}$ &-0.0360 & $t_{32}^{(1,1,1)}$ & 0.0001 \\
 $t_{11}^{(1,1,0)}$ & 0.0894 & $t_{22}^{(1,1,0)}$ &-0.0606 &                    &        &&  &$t_{11}^{(1,1,0)}$ & 0.0043 & & $t_{22}^{(1,1,0)}$ & 0.0105 & & $t_{33}^{(1,1,0)}$ &-0.0497 &                    &        \\
 $t_{11}^{(1,0,1)}$ & 0.0000 & $t_{22}^{(1,0,1)}$ & 0.1988 &                    &        &&  &$t_{11}^{(1,0,1)}$ &-0.1338 & & $t_{22}^{(1,0,1)}$ & 0.0000 & & $t_{33}^{(1,0,1)}$ & 0.0105 &                    &        \\
 $t_{11}^{(1,1,1)}$ & 0.0134 & $t_{22}^{(1,1,1)}$ & 0.0281 &                    &        &&  &$t_{11}^{(1,1,1)}$ & 0.0038 & & $t_{22}^{(1,1,1)}$ & 0.0018 & & $t_{33}^{(1,1,1)}$ &-0.0113 &                    &        \\
\hline
\hline
\end{tabular}
\end{table*}

\subsection{Analysis of EBRs and Orbitals}
By doing the analysis of the elementary band representations (EBR) in the theory of topological quantum chemistry\cite{bradlyn2017topological}, one can easily obtain orbital information in real space. An EBR of $\rho@q$ is labeled by the Wyckoff position $q$ and the irrep $\rho$ of its site symmetry group. At first, the irreps of the six low-energy bands at maximal high-symmetry $k$-points~\cite{gao2020,vergniory2019complete} are computed without SOC. The results are listed in Table.~\ref{tab:irreps}. In the crystal of LaNiO$_2$, the Ni atom sites at Wyckoff position $1a [0,0,0]$, while La atom sites at Wyckoff position $1d [0.5,0.5,0.5]$. Both Wyckoff positions have the site-symmetry group of $4/mmm$.  Note that five $d$-orbitals only support the irreps of $A_{1g}~(d_{3z^2-r^2})$, $B_{1g}~(d_{x^2-y^2})$, $B_{2g}~(d_{xy})$, and  $E_g~(d_{xz,yz})$ under the single group of $4/mmm$.

The results of the EBR analysis on atomic orbitals are addressed as follows.
First, by switching the irreps of $A3-$ and $A4+$ at A, we can find that the four occupied bands can be represented as the sum of three EBRs: $A_{1g}@1a~\oplus~B_{2g}@1a~\oplus~E_{g}@1a$. Among them, the $3d_{xy}$-based EBR $B_{2g}@1a$ has highest states at A, which may intersect with the chemical potential in the hole-doping case.
Second, the band of $3d_{x^2-y^2}$-induced EBR $B_{1g}@1a$ is clear shown by the weights in Fig.~\ref{fig:2}(a). There is no doubt that $3d_{x^2-y^2}$ orbital contributes the largest Fermi surface.
Third, the irrep GM$1+$ at $\Gamma$ is from the $5d_{3z^2-r^2}$-induced EBR $A_{1g}@1d$.
Lastly, the inverted irrep $A3-$ is from the $5d_{xy}$-induced ERB $B_{2g}@1d$. 
So far, all of the orbital compounds are consistent with our DFT calculations in Fig.~\ref{fig:2}(a).

Then, we aim to construct a minimal effective model to reproduce the bands and symmetry characteristics near $E_F$. Besides the $3d_{x^2-y^2}$-induced EBR of $B_{1g}@1a$, the two irreps of A$3-$ and GM$1+$ can be generated in the EBR of $A_{1g}@1b$. Therefore, we derive a two-band model, consisting of two EBRs: $B_{1g}@1a~\oplus~A_{1g}@1b$, which reproduce the exact irreps of symmetries near $E_F$ from the DFT calculations. Even through there is no actually atomic orbital at Wyckoff position $1b$ [0,0,0.5] (with the site symmetry group $4/mmm$), it could be formed by the hybridization of the atomic orbitals on other sites. The two-band model would be constructed for the un-doped case in the following section.

\subsection{Two-band effective model}
Under the basis of the $B_{1g}$ orbital at the $1a$ Wyckoff position and $A_{1g}$ orbital at the $1b$ Wyckoff position, the tight binding (TB) model is constructed as follows. The diagonal terms in the Hamiltonian are 
\begin{equation}
  \begin{split}
    T_{\alpha\alpha} &= t_{\alpha\alpha}^{(0,0,0)} + 2t_{\alpha\alpha}^{(1,0,0)}(\cos(k_x)+\cos(k_y)) \\
                     & + 2t_{\alpha\alpha}^{(0,0,1)}\cos(k_z)+ 4t_{\alpha\alpha}^{(1,1,0)}\cos(k_x)\cos(k_y) \\ 
                     &+ 4t_{\alpha\alpha}^{(1,0,1)}\cos(k_z)(\cos(k_x)+\cos(k_y)) \\ 
                     &+ 8t_{\alpha\alpha}^{(1,1,1)}\cos(k_x)\cos(k_y)\cos(k_z) 
\end{split}
\end{equation}
where $\alpha = {1,2}$ represent the $B_{1g}$@$1a$ and the $A_{1g}$@$1b$, respectively. $t_{\beta\alpha}^{(l,m,n)}$ stands for the hopping amplitude from orbital $\beta$ of the original cell to $\alpha$ of the $(l,m,n)$ cell:
\begin{equation}
  t_{\beta\alpha}^{(l,m,n)} \equiv \langle \beta;000 | \hat H | \alpha;lmn \rangle 
\end{equation}
In the off-diagonal term, the nearest and next-nearest hoppings between different orbitals are given as, 
\begin{equation}
\begin{split}
 S = t_{21}^{(1,0,0)} (1+e^{ik_z})(4\cos(k_x) - 4\cos(k_y))
\end{split}
\end{equation}
Thus, our two-band model Hamiltonian is written as 
\begin{equation}
  H_2(k) = \begin{pmatrix}
    T_{11} & \dagger \\ 
    S & T_{22} \\
  \end{pmatrix}
\end{equation}
The fitting results are shown in the Fig.~\ref{fig:2}(f) and the parameters can be found in Table \ref{tab:parawzj}.
This two-band model can reproduce all the Fermi surfaces and symmetry characteristics obtained from the DFT calculations.

\begin{figure}[!t]
\centering
\includegraphics[width=8.5cm]{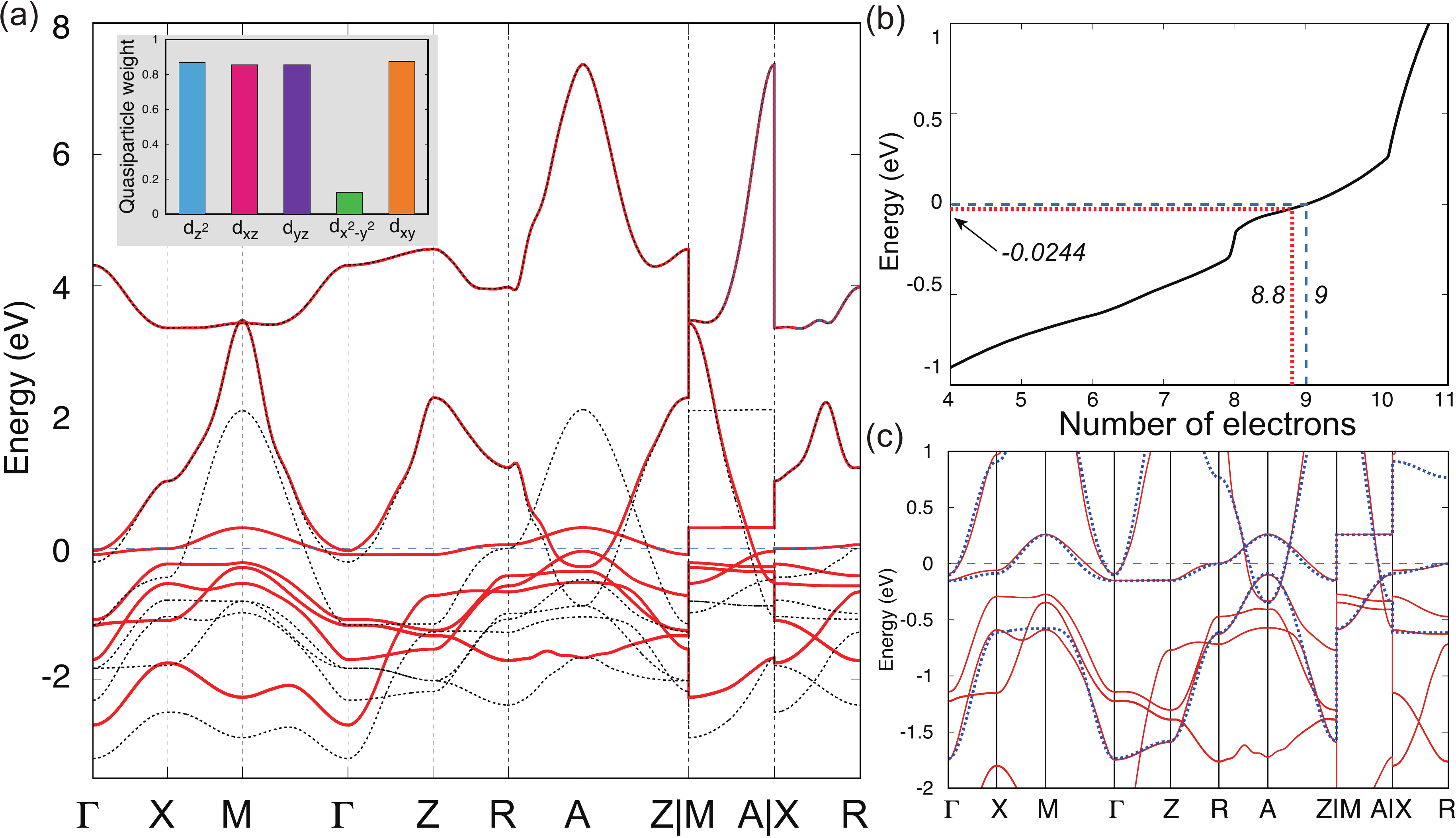}
\caption{(Color online)
(a) The band structure of the non-interacting tight-banding Hamiltonian $H_{tb}$, extracted by the $Wannier90$ package, is plotted in the black dashed lines, while the LDA+Gutzwiller bands are plotted in red solid lines. The inset of (a) shows the quasiparticle weights of five $3d$ orbitals. (b) shows the total number of electrons as a function of chemical potential in the LDA+Gutzwiller band structure. In panel (c), the bands of the three-band model are presented in blue dashed lines. 
}
\label{fig:3}
\end{figure}

\subsection{LDA+Gutzwiller method}
To treat the correlation effect of five Ni $3d$ orbitals, we have employed the LDA+Gutzwiller method~\cite{xydeng,ldu}. The corresponding Gutzwiller trial wave function has been constructed as $\ket{G} = \hat{P}\ket{0}$ with
\begin{equation}
\hat{P} = \prod_i \hat{P}_i = \prod_i \sum_{\Gamma \Gamma '} \lambda_{i;\Gamma\Gamma'} \ket{i,\Gamma}\bra{i,\Gamma'}
\end{equation} 
where $\ket{0} $ is the noninteracting wave function and $\hat{P}$ is the Gutzwiller local projector with  $\ket{i,\Gamma}$ the atomic eigenvectors on site $i$. $ \lambda_{i;\Gamma\Gamma'}$ is the so-called Gutzwiller variational parameters which adjusts the weight of different local atomic configurations. Ground states are obtained by minimizing the total energy $E=\bra{G}(H_{tb}+H_{dc}+H_{int})\ket{G}$ with some Gutzwiller constraints. The non-interacting Hamiltonian ($H_{tb}$) is extracted by the $Wannier90$ package~\cite{wannier90rmp} from the DFT calculations without SOC, which contains two La $5d_{3z^2-r^2,xy}$ orbitals and five Ni $3d$ orbitals. The double-counting term $H_{dc}$ is given self-consistently. The on-site interacting term takes the Slater-Kanamori rotationally invariant atomic interaction~\cite{nicola}:
\begin{equation}
\begin{split}
H_{int}&=U\sum_\alpha \hat n_{a\up}\hat n_{a\dw}+\frac{U'}{2}\sum_{a\neq b}\sum_{\sigma\sigma'} \hat n_{a\sigma}\hat n_{b\sigma'} \\
&-\frac{J}{2}\sum_{a\neq b}\sum_{\sigma} c^\dagger_{a\sigma}c_{a-\sigma}c^\dagger_{b-\sigma}c_{b\sigma} \\
&-\frac{J'}{2}\sum_{a\neq b}             c^\dagger_{a\up}c^\dagger_{a\dw}c_{b\up}c_{b\dw}
\end{split}
\end{equation}
where $c^\dagger_{a\sigma}~(c_{a\sigma})$ creates (annihilates) an electron in the state of the orbital $a$ and the spin $\sigma$, and $\hat n_{a\sigma}=c^\dagger_{a\sigma} c_{a\sigma}$.

 For simplicity, we adopt diagonal variational parameters which means $\lambda_{i;\Gamma\Gamma'} = \lambda_{i;\Gamma\Gamma} \delta_{\Gamma\Gamma'}$ here. Five Ni $3d$ orbitals are treated as correlated orbitals in the calculation (which correspond to 10 bands once considering the spin degree of freedom). 
%The onsite electron interaction is formulated by the Kanamori rotationally invariant atomic interaction and 
We take Coulomb interaction $U$ of 5 eV, Hund's coupling $J$ of $0.18U$, $U'=U-2J$ and $J'=J$. Besides, the occupancy of Ni $3d$ orbitals has been forced to be 8.462 obtained from the DFT calculations. 
The results show that the quasi-particle weight of $d_{x^2-y^2}$ orbital is very small, $0.12$, while the weights of other four orbitals are about $0.85$. 
After considering the renormalization of the correlated $3d$ orbitals, the modified band structure is obtained, as shown in Fig.~\ref{fig:3}(a).
Two significant features are found after our Gutzwiller correction: one is that the bandwidth of $3d_{x^2-y^2}$ has been renormalized largely, leading to a DOS peak around Fermi level, which may contribute to the large peak around zero energy in RIXS spectrum\cite{Hepting2020electronic}; the other is that the $3d_{xy}$ states near A point become very close to $E_F$. As a result, a HP could be induced by hole doping, which may be related to the observed sign change of Hall coefficient. 

In addition, the band inversion between the $3d_{xy}$ states and $5d_{xy}$ states near A point could be important in the hole-doped nickelate compound Nd$_{0.8}$Sr$_{0.2}$NiO$_2$, since they are very close to $E_F$. Therefore, we modify our model by simply adding an additional $3d_{xy}$ orbital to capture the potential band inversion in this hole-doped compound. The modified model is written as,
% \begin{equation}
%   \begin{split}
%   H_3(k)& = \begin{pmatrix}
%     T_{11} &           &   \dagger \\
%     S & T_{22} &           \\ 
%     W & P & T_{33} \\
%   \end{pmatrix} \\
%  \text{ with }W& = 8t_{31}^{(1,1,1)}\cos(k_x)\cos(k_y)\cos(k_z) \\
%   P &= 4t_{32}^{(1,0,0)}(1+e^{1k_z})(\cos(k_x)-\cos(k_y))
%   \end{split}
% \end{equation}
\begin{equation}
  \begin{split}
  H_3(k)& = \begin{pmatrix}
    T_{11} &           &   \dagger \\
    S & T_{22} &           \\ 
    W & P & T_{33} \\
  \end{pmatrix} \\
 \text{ with }P &= 4t_{31}^{(1,0,0)}(1+e^{1k_z})(\cos(k_x)-\cos(k_y)) \\
   W& = 8t_{32}^{(1,1,1)}\cos(k_x)\cos(k_y)\cos(k_z)
  \end{split}
\end{equation}
The parameters are obtained by fitting with the renormalized bands, as shown in Table \ref{tab:parawzj}. The results of the modified model $H_3(k)$ are shown in Fig.~\ref{fig:3}(c), which fit very well with the LDA+Gutzwiller bands, which can be compared with the ARPES experimental data.
In any case, a four-band model constructed totally from the real atomic orbitals are presented in the SM.

\begin{figure}[!t]
\centering
\includegraphics[width=8cm]{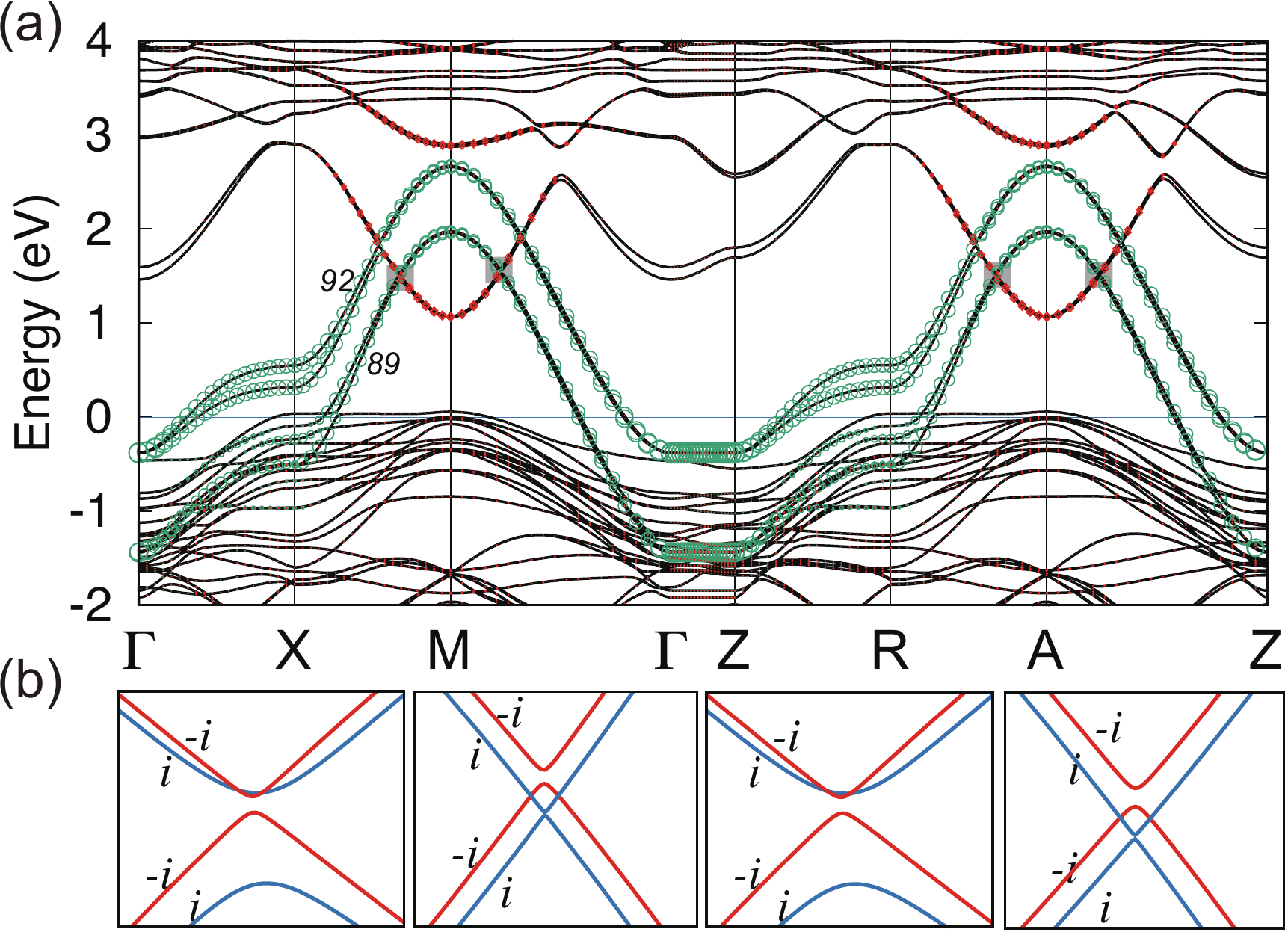}
\caption{(Color online)
(a) The ferromagnetic band structure of La$_3$Ni$_2$O$_6$ with SOC. The weights of La $5d_{xy}$ and Ni $3d_{x^2-y^2}$ states are indicated by the size of red diamonds and green circles, respectively.
(b) The crossing points between 90 and 91 bands. Bands with different mirror eigenvalues are printed as blue (for $+i$) and red (for $-i$) lines.
}
\label{fig:4}
\end{figure}

\subsection{Ferromagnetic state in La$_3$Ni$_2$O$_6$}
Band topology can also be found in another member of the $T^\prime$ type $Ln$$_{n+1}$Ni$_{n}$O$_{2n+2}$ homologous series, for example the ferromagnetic state of La$_3$Ni$_2$O$_6$ ($n=2$), whose band structure with SOC is shown in Fig.~\ref{fig:4}(a). The $z$-oriented magnetism reduces the symmetry from type-II magnetic SG 139 to type-I magnetic SG 87. 
Near the M/A point, the four downward parabolic bands(\ie 89-92 bands) are mainly from the $3d_{x^2-y^2}$ states of Ni atoms in two planar Ni--O layers, while the two upward parabolic bands (\ie 93-94 bands) are mainly from the $5d_{xy}$ states of the La atoms sandwiching by the two Ni--O planes [Fig.~\ref{fig:1}(b)].
Once looking closely at the crossings between 90th and 91st bands in Fig.~\ref{fig:4}(b), we found that there is a gap along X--M (R--A), while there are two crossing points along M--$\Gamma$ (A--Z). These crossing points are parts of the tiny $M_z$-protected nodal rings at $k_z=0,2\pi/c$ planes, which can be easily removed by very small perturbations (without changing the ordering of energy bands at high-symmetry $k$-points).
According to symmetry indicators defined in magnetic systems\cite{PhysRevB.85.165120}, insulators in type-I MSG 87 is characterized by the indicators $\mathbb{Z}_4 \times \mathbb{Z}_4$ \cite{po2017symmetry,ono2018unified}, which can be explained by two mirror Chern numbers in the $k_z = 0$ plane.
By using the C$_4$ eigenvalues\cite{fang2012bulk}, we find that the Chern numbers are $0,2$ for the mirror eigenvalue $\pm i$ sectors, respectively, which indicate that non-trivial edge states can emerge on the $M_z$-preserving surface.

% The crossing between these bands along M$\Gamma$ and AZ are protected by $m_z$ symmetry, which are still robust when including SOC because the crossing are happened between different spin channel. Thus we can conclude that there are two nodal loops on the $k_z=0$ and $k_z=\pi$ plane, respectively.

% Without SOC, the crossings between these bands are protected by $m_z$ symmetry, in each spin channel.
% Once including SOC, they are all gapped out, leading that there is always a direct gap between the 90th band and 91st band.
% According to symmetry indicators defined in magnetic systems\cite{PhysRevB.85.165120}, 
% type-I MSG 87 is characterized by the indicators $\mathbb{Z}_4 \times \mathbb{Z}_4$ \cite{po2017symmetry,ono2018unified}, which can be explained by the mirror Chern number in the $k_z = 0$ plane. By using the C$_4$ eigenvalues\cite{fang2012bulk}, we find that the Chern numbers are 0,2 for the mirror eigenvalue $\pm i$ sectors, respectively, which indicate that non-trivial edge states can emerge on the $m_{001}$-preserving surface. 
%Details of the calculation of mirror Chern numbers can be seen in the SM, in which we also discuss some results with different filling number.
%Note that the symmetry eigenvalues method can only diagnose the mirror Chern number module 4, thus we can not decide the number of edge states from that. 

\section{Conclusion}
Based on our first-principles calculations, 
we find that, in the un-doped case, there are three EPs: a largest EP from Ni $3d_{x^2-y^2}$ orbital, a small one at $\Gamma$ from La $5d_{3z^2-r^2}$ orbital and a relatively bigger one at A from La $5d_{xy}$.  These Fermi surfaces and symmetry characteristics can be reproduced by our two-band model, which consists of two elementary band representations: $B_{1g}@1a~\oplus~A_{1g}@1b$. 
In the obtained band structure, there is a band inversion happened near A, which gives rise to a pair of Dirac points along A--M once including SOC. The correlation effect of Ni $3d$ orbitals have been estimated in our LDA+Gutzwiller calculation, and the renormalized band structure is obtained, which indicates that Ni $3d_{xy}$ states become very close to $E_F$ near A point. A hole pocket is likely induced by hole doping, which may be related to the observed sign change of Hall coefficient. 
The hole-pocket band and the band inversion can be captured in the modified model by simply including another Ni $3d_{xy}$ orbital. In addition, we show that the nontrivial band topology in the ferromagnetic two-layer compound La$_3$Ni$_2$O$_6$ is associated with Ni $3d_{x^2-y^2}$ and La $5d_{xy}$ orbitals.

\noindent \textbf{Note added.} 
At the stage of finalising the present paper, we are aware of the similar works on nickelates\cite{hirsch2019hole,sakakibara2019model,botana2019similarities,xianxinwu2019robust,yusuke2019formation}.

\ \\
\noindent \textbf{Acknowledgments}
We thank Jinguang Cheng for drawing our attention to this subject and valuable discussion, as well as critical reading of this work.
This work was supported by the National Natural Science Foundation of China (11974395, 11504117, 11774399, 11622435, U1832202), Beijing Natural Science Foundation (Z180008), the Ministry of Science and Technology of China (2016YFA0300600, 2016YFA0401000 and 2018YFA0305700), the Chinese Academy of Sciences (XDB28000000, XDB07000000), the Beijing Municipal Science and Technology Commission (Z181100004218001, Z171100002017018). H.W. acknowledges support from the Science Challenge Project (No.~TZ2016004), the K. C. Wong Education Foundation (GJTD-2018-01). 
Z.W. acknowledges support from the CAS Pioneer Hundred Talents Program and the National Thousand-Young-Talents Program.

\bibliography{tps}

% \end{document}
\clearpage
\begin{widetext}
\beginsupplement{}
\section*{SUPPLEMENTARY MATERIAL}
\subsection*{A. The band structures of different compounds $Ln$NiO$_2$ ($Ln$=La, Pr, Nd) }
The band structures are in our DFT calculations with SOC for the $Ln$NiO$_2$ ($Ln$=La, Pr, Nd) compounds. 
\begin{figure*}[!hb]
\centering
\includegraphics[width=17cm]{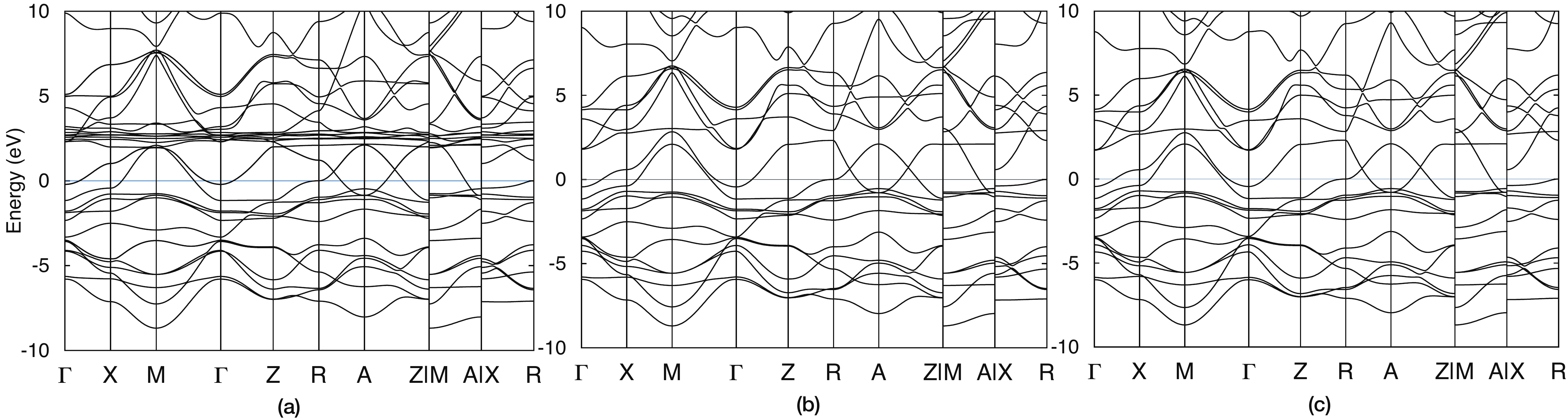}
\caption{(Color online)
Band structures of LaNiO$_2$ (a), PrNiO$_2$ (b) and NdNiO$_2$ (c). The $4f$ states of Pr and Nd treated as core states in our PAW potential in our VASP calculations.
}
\label{fig:supp1}
\end{figure*}

\subsection*{B. The evolution of band structures with different U values}
The band structure can be affected by the Coulomb interaction, which can be simulated by using the LDA+U method as implemented in VASP. Here we add the on-site interaction on Ni-$3d$ orbitals from $U=1$eV to $U=6$eV in the LaNiO$_2$ compound. The interaction does not change the overall band structure very much. 
But, the relative energy difference between the bands of four $3d$ orbitals (except $d_{x^2-y^2}$) and the other bands (\ie $3d_{x^2-y^2},~5d_{xy}$) increases monotonically as increasing U.
As shown in Fig.~\ref{fig:supp2}, the band inversion can be removed when $U=6$eV, resulting the disappearance of Dirac points.

\begin{figure*}[!hb]
\centering
\includegraphics[width=15cm]{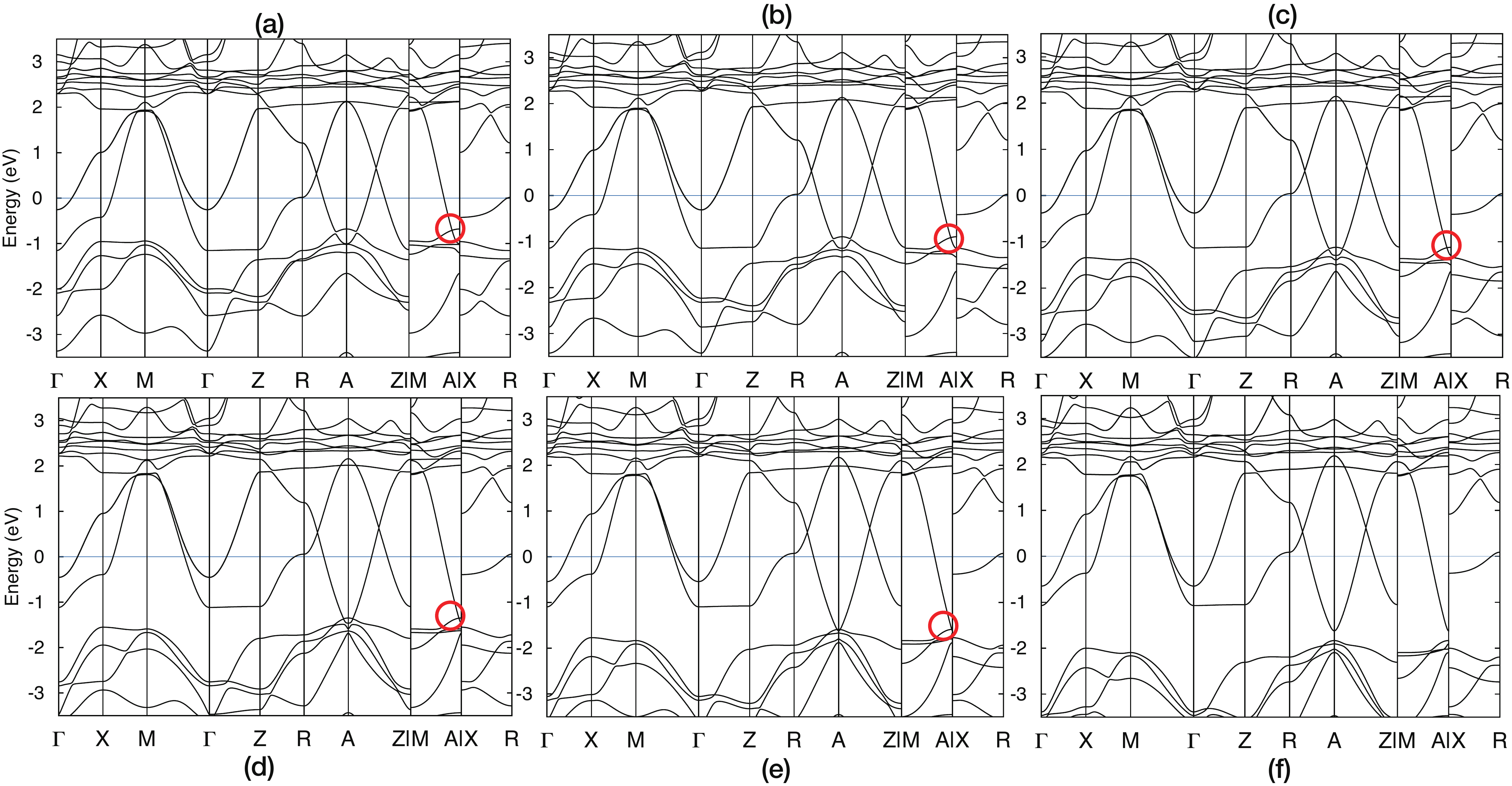}
\caption{(Color online)
Band structure of LaNiO$_2$ under different Coulomb interactions, from $U=1$eV (a) to $U=6$eV (f). The Dirac point concerned are indicated by red circles.
}
\label{fig:supp2}
\end{figure*}

\subsection*{C. Irreducible representions for DFT bands with SOC}
The irreps of the energy bands with SOC at maximal high-symmetry points are give in Table \ref{tab:irrepso}. By using the opensource codes of topological quantum chemistry~\cite{gao2020,vergniory2019complete,bradlyn2017topological}, the results indicate that it's a Dirac semimetal with 8 occupied bands.
\begin{table}[!h]
\caption{
The irreducible representations (irreps) for the eight occupied bands (4 doubly-degenerate ones) with SOC at the maximal high-symmetry points in SG 123. The irreps are given in ascending energy order. The notation of $Zm(n)$ implies the irrep $m$ at the $Z$ point with the degeneracy of $n$.
}\label{tab:irrepso}
\begin{tabular}{cccccc}
\hline
\hline
 A & $\Gamma$ & M & R & X & Z \\
\hline
 A9(2) & GM7(2) & M7(2) & Z6(2) & R5(2) & X5(2) \\
 A7(2) & GM6(2) & M6(2) & Z7(2) & R5(2) & X5(2) \\
 A7(2) & GM7(2) & M7(2) & Z6(2) & R5(2) & X5(2) \\
 A6(2) & GM6(2) & M6(2) & Z7(2) & R5(2) & X5(2) \\
\hline
\hline
\end{tabular}
\end{table}

\subsection*{D. The four-band model}
Here we construct a four band TB model to capture the band topology and Fermi surface structure by using the real atoms' orbitals, in case anyone needs the real atomic orbitals to determine the interacting strength.
Based on the analysis of orbital components, we choose our orbitals as: {La-$d_{xy}$, La-$d_{z^2}$, Ni-$d_{x^2-y^2}$, Ni-$d_{xy}$}, which are referred to as $\alpha=1,2,3,4$, respectively.
The four-band model reads as:
\begin{equation}
  H_4(k) = \begin{pmatrix}
    H_{11} &   &        &   \dagger \\
    H_{21} & H_{22} & &       \\ 
    H_{31} & H_{32} & H_{33}& \\
    H_{41} & H_{42} & H_{43}& H_{44} \\
  \end{pmatrix} 
\end{equation}
The space group $\mathcal{G} $ is $P4/mmm$, generated by discrete translation symmetry, inversion, $C_{4z}$, $C_{2x}$ and $C_{2y}$.
\\
\begin{equation}
  t_{\beta\alpha}^{(l,m,n)} \equiv \langle \beta;000 | \hat H | \alpha;lmn \rangle 
\end{equation}
\indent The diagonal terms have the same form as Eq~(1) in the main text
\begin{equation}
  \begin{split}
    H_{\alpha\alpha} &= t_{\alpha\alpha}^{(0,0,0)} + 2t_{\alpha\alpha}^{(1,0,0)}(\cos(k_x)+\cos(k_y)) \\
                     & + 2t_{\alpha\alpha}^{(0,0,1)}\cos(k_z)+ 4t_{\alpha\alpha}^{(1,1,0)}\cos(k_x)\cos(k_y) \\ 
                     &+ 4t_{\alpha\alpha}^{(1,0,1)}\cos(k_z)(\cos(k_x)+\cos(k_y)) \\ 
                     &+ 8t_{\alpha\alpha}^{(1,1,1)}\cos(k_x)\cos(k_y)\cos(k_z) 
\end{split}
\end{equation}
where $t_{\beta\alpha}^{(l,m,n)}$ stands for the hopping amplitude from orbital $\beta$ of the original cell to $\alpha$ of the $(l,m,n)$ cell as shown in Fig.~\ref{fig:supp3}(a):
\\
\indent The anti-diagonal terms are listed as follows
\begin{equation}
  H_{21} = 2t_{21}^{(1,1,0)}(\cos(k_x+k_y)-\cos(k_x-k_y))
\end{equation}
\begin{equation}
  H_{31} = t_{31}^{(1,0,0)}(1+e^{ik_z})[(1-e^{ik_y})(e^{-ik_x}-e^{2ik_x})+(1-e^{ik_x})(e^{2ik_y}-e^{-ik_y})]
\end{equation}
\begin{equation}
  H_{32} = t_{32}^{(1,0,0)}(1+e^{ik_z})[(1+e^{ik_y})(e^{2ik_x}+e^{-ik_x})-(1+e^{ik_x})(e^{2ik_y}+e^{-ik_y})]
\end{equation}
% \begin{equation}
%   \begin{split}
%   H_{41} &= t_{41}^{(0,0,0)}(1+e^{ik_z})(1+e^{-ik_x})(1+e^{-ik_y}) \\ 
%          &+ t_{41}^{(1,0,0)}(1+e^{ik_z})[(1+e^{ik_y})(e^{2ik_x}+e^{-ik_x})+(1+e^{ik_x})(e^{2ik_y}+e^{-ik_y})]
%   \end{split}
% \end{equation}
\begin{equation}
  H_{41} = t_{41}^{(0,0,0)}(1+e^{ik_z})(1+e^{ik_x})(1+e^{ik_y})
\end{equation}
\begin{equation}
  H_{42} = t_{42}^{(0,0,0)}(1+e^{ik_z})(1-e^{ik_x})(1-e^{ik_y})
\end{equation}
\begin{equation}
  H_{43} = 8t_{43}^{(1,1,1)}\cos(k_x)\cos(k_y)\cos(k_z)
\end{equation}
Using the parameters listed in Table.~\ref{tab:para4band}, the band structure of the four band model are shown in Fig.~\ref{fig:supp3}(b).

\begin{figure*}[!tb]
\centering
\includegraphics[width=14cm]{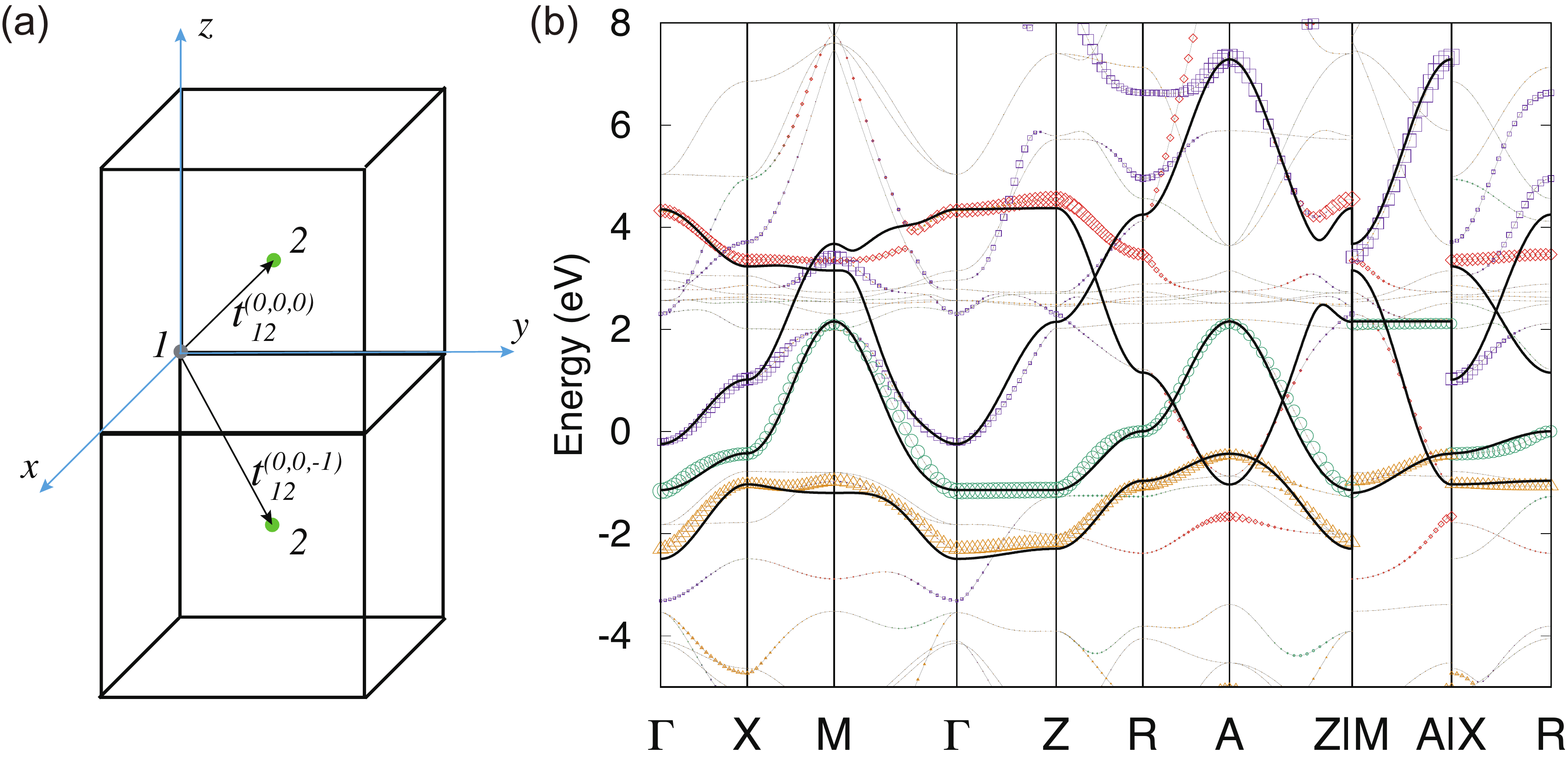}
\caption{(Color online)
(a) Hopping between different orbitals
(b) Band structure of four band tight binding model (black solid line).
}
\label{fig:supp3}
\end{figure*}

\begin{table*}[t]
  \caption{
  The hopping parameters for the four-band Hamiltonian
  }\label{tab:para4band}
  \begin{tabular}{cc|cc|cc|cc|cc}
  \hline
  \hline
   $t_{11}^{(0,0,0)}$ & 2.4319 & $t_{22}^{(0,0,0)}$ & 2.8954 & $t_{33}^{(0,0,0)}$ & 0.1470 & $t_{44}^{(0,0,0)}$ &-1.2546 & $t_{21}^{(1,1,0)}$ & 0.2720 \\
   $t_{11}^{(1,0,0)}$ & 0.4031 & $t_{22}^{(1,0,0)}$ &-0.5519 & $t_{33}^{(1,0,0)}$ &-0.4125 & $t_{44}^{(1,0,0)}$ &-0.2013 & $t_{31}^{(1,0,0)}$ &-0.0281 \\
   $t_{11}^{(0,0,1)}$ & 0.5106 & $t_{22}^{(0,0,1)}$ &-0.7931 & $t_{33}^{(0,0,1)}$ &-0.0536 & $t_{44}^{(0,0,1)}$ &-0.0450 & $t_{32}^{(1,0,0)}$ & 0.0281 \\
   $t_{11}^{(1,1,0)}$ & 0.0597 & $t_{22}^{(1,1,0)}$ & 0.0659 & $t_{33}^{(1,1,0)}$ & 0.0894 & $t_{44}^{(1,1,0)}$ &-0.0638 & $t_{41}^{(0,0,0)}$ &-0.1286 \\
   $t_{11}^{(1,0,1)}$ &-0.1366 & $t_{22}^{(1,0,1)}$ & 0.0453 & $t_{33}^{(1,0,1)}$ & 0.0000 & $t_{44}^{(1,0,1)}$ & 0.0156 & $t_{42}^{(0,0,0)}$ & 0.1366 \\
   $t_{11}^{(1,1,1)}$ &-0.0023 & $t_{22}^{(1,1,1)}$ & 0.0036 & $t_{33}^{(1,1,1)}$ & 0.0134 & $t_{44}^{(1,1,1)}$ &-0.0069 & $t_{43}^{(1,1,1)}$ & 0.0001 \\
  \hline
  \hline
  \end{tabular}
  \end{table*}

\newpage

\end{widetext}
\end{document}